\documentclass[iop,revtex4]{emulateapj}
\usepackage{amsmath}
\usepackage{color}

\begin{document}

\slugcomment{submitted to ApJ}
\shorttitle{ROTATION OF HOT GAS AROUND THE MILKY WAY}
\shortauthors{HODGES-KLUCK, MILLER, AND BREGMAN}

\title{The Rotation of the Hot Gas Around the Milky Way}

\author{Edmund J. Hodges-Kluck$^{1}$, Matthew~J. Miller$^{1}$ \&
  Joel~N. Bregman$^{1}$}
\altaffiltext{1}{Department of Astronomy, University of Michigan, Ann
  Arbor, MI 48109}
\email{hodgeskl@umich.edu}

\begin{abstract}
The hot gaseous halos of galaxies likely contain a large amount of mass
and are an integral part of galaxy formation and evolution. The Milky Way 
has a $2 \times 10^6$~K halo that is detected in emission and by absorption 
in the \ion{O}{7} resonance line against bright background AGNs, 
and for which
the best current model is an extended spherical distribution.  
Using
XMM-Newton RGS data, we measure the Doppler shifts of the \ion{O}{7} absorption-line 
centroids toward an ensemble of AGNs.  These Doppler shifts  
constrain the dynamics of the hot halo, 
ruling out a stationary halo at about
$3\sigma$ and a co-rotating halo at $2\sigma$, 
and leading to a best-fit rotational velocity $v_{\phi} = 183\pm 41$\,km\,s$^{-1}$
for an extended halo model. 
These results suggest that the hot gas rotates and that it contains
an amount of angular momentum comparable to that in the stellar disk.
We examined the possibility of a model with a kinematically distinct disk and
spherical halo. To be consistent with the emission-line X-ray data the disk
must contribute less than 10\% of the column density, implying that the
Doppler shifts probe motion in the extended hot halo.
\end{abstract}


\section{Introduction}

A basic prediction of $\Lambda$CDM galaxy-formation models is the existence
of a hot ($10^6-10^7$\,K) halo of gas accreted from the intergalactic medium
around Milky Way-sized galaxies (extending to the virial radius),
which forms as infalling gas is heated to the virial temperature at an
accretion shock \citep[e.g.,][]{white91}. These halos may provide most of the
fuel for long-term star formation in these galaxies \citep{crain10,joung12},
but their predicted properties are sensitive to the input physics, which
can be constrained by the measurable properties of the gas.

Based on work over the past several years, we know that these extended halos
exist, including around the Milky Way \citep{anderson11,anderson13,miller15}.
The extent and luminosity of the hot gas implies that it has a similar mass to
the stellar disk, and therefore could play an important role in galaxy
evolution. Thus, it is important to measure the properties of the hot gas
beyond mass and temperature (such as metallicity and density or velocity
structure). However, hot halos are faint
and the measurable X-ray luminosity can be dominated by stellar feedback
ejecta near the disk \citep{li13}, which makes these measurements difficult.

Only in the Milky Way can one
measure the structure, temperature, metallicity, and kinematics of the
hot gas through emission and absorption lines
\citep{nicastro02,paerels03,mckernan04,yao05,williams05,fang06,williams06,williams07,yao09,henley12,gupta12},
but kinematic constraints from prior studies \citep{bregman07,fang15}
are weak. Recent developments in the calibration of the X-ray grating
spectrometers and the accumulation of multiple high quality data sets for
individual objects have made it possible
to determine line centroids to an accuracy of tens of km\,s$^{-1}$, which
enables us to improve the constraints on the kinematics of the gas by
measuring Doppler shifts in lines that trace the hot gas. 
The $\lambda$21.602\AA\ resonance \ion{O}{7} absorption line \citep{drake88}
is the best candidate, 
since it is sensitive to temperatures of $10^{5.5}-10^{6.3}$\,K (which includes
much of the Galactic coronal gas)
and it is detected at zero redshift towards a large number of background 
continuum sources \citep[e.g.,][]{fang15}. The \textit{emission} lines 
produced by the same species are useful for determining the structure
and temperature of the hot gas \citep{henley12,miller15}, but they are
too faint for high resolution spectroscopy. In this paper we constrain,
for the first time, the radial and azimuthal velocity of the hot gas by
measuring the Doppler shifts in \ion{O}{7} lines detected towards bright
sources outside the disk of the Milky Way.

\section{Observations and Data Analysis}

\subsection{Sample and Reduction}

To measure the global velocity of the million-degree gas around the Galaxy, 
one needs to measure Doppler shifts towards a range of sources across the sky 
in lines sensitive to this temperature. This gas is detected in X-ray emission and
absorption, but the emission lines are far too faint for a focused grating
observation. X-ray imaging CCDs measure the energies of incoming photons
and are thus also low resolution spectrometers, but their spectral resolution
is far too low to measure Doppler shifts of tens of km\,s$^{-1}$. The only
instruments capable of this accuracy are the \textit{Chandra} Low/High-Energy
Transmission Grating (LETG/HETG) Spectrometers and the \textit{XMM-Newton}
Reflection Grating Spectrometer (RGS), and the 21.602\AA\ \ion{O}{7} line is the 
only line that probes the relevant temperatures and is detected at $z=0$ 
towards many background continuum sources.
The O{\sc viii} line at 18.96\AA\ probes slightly hotter gas and is only
detected towards a few objects. 

\begin{figure*}[t]
\begin{center}
\hspace{-0.4cm}\includegraphics[width=0.5\textwidth]{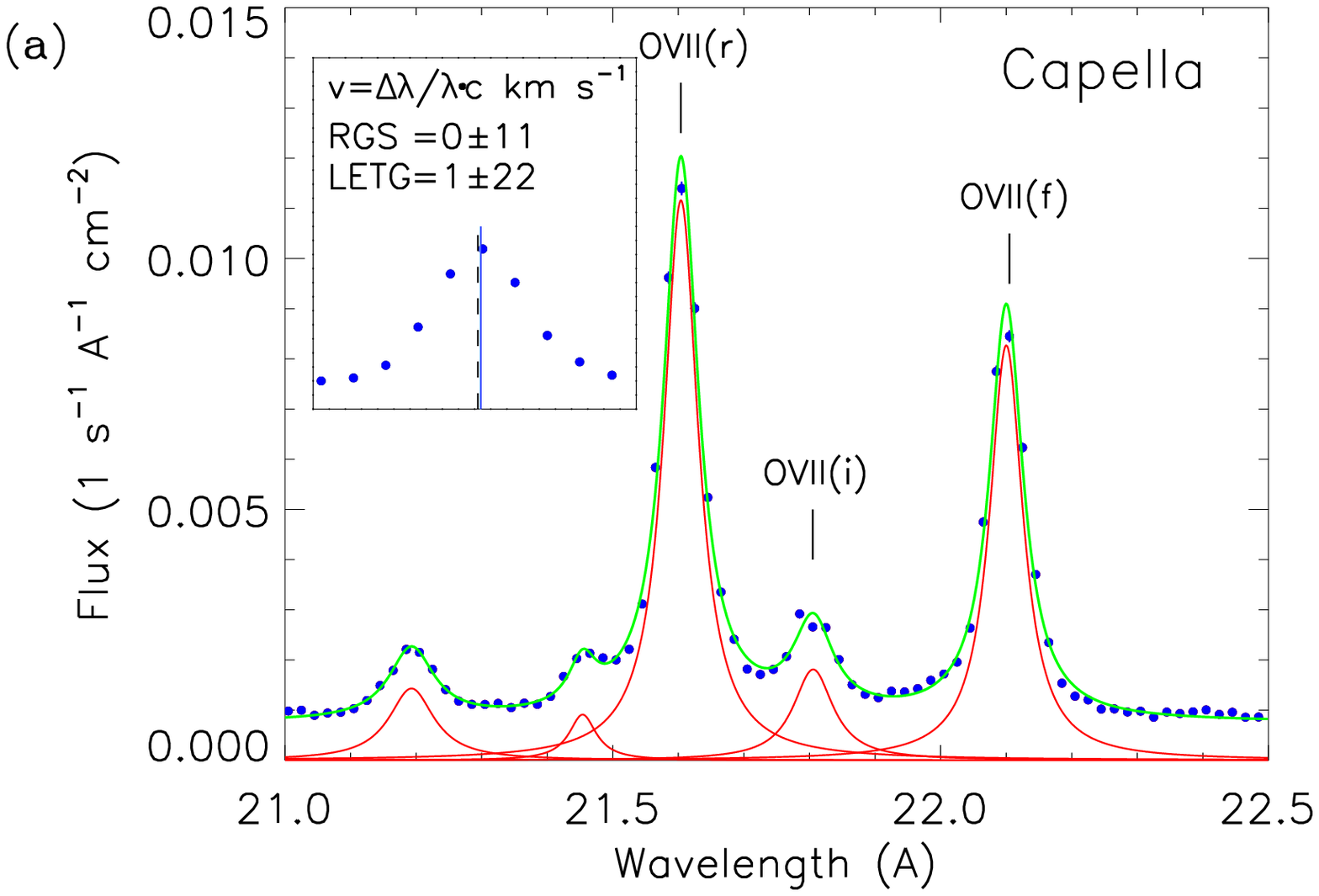}
\hspace{-0.4cm}\includegraphics[width=0.5\textwidth]{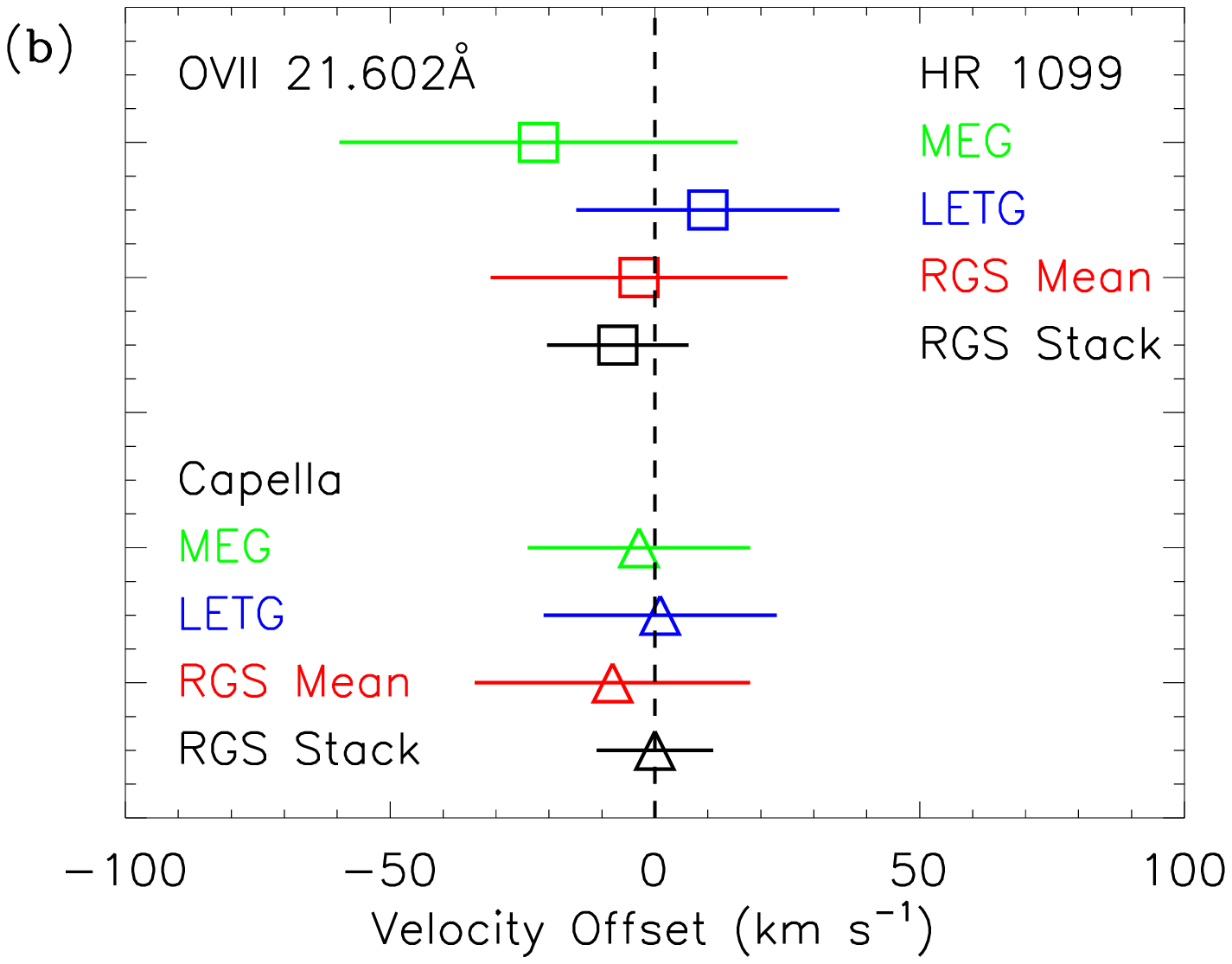}
\caption{(a) The stacked RGS spectrum for Capella shows
  that the reduction protocols achieve an accurate wavelength solution
  (inset), and that averaging multiple data sets substantially reduces
  the systematic scatter in the wavelength grid. (b) Our measurements of
  the \ion{O}{7} emission line centroid relative to the
  21.602\AA\ rest wavelength in three instruments and two stars show
  that there are no serious cross-instrument issues with the
  wavelength grid. In each case, we accounted for the radial velocity
  of the stars.
}
\label{figure.capella_wavelength}
\end{center}
\end{figure*}

Our initial sample included all archival LETG and RGS data sets where the
line has been detected in the literature. The LETG has modestly better
spectral resolution at 21.6\AA\ ($R=400$) than the RGS ($R=325$), but it
only has a third of the effective area at this wavelength 
(15\,cm$^{2}$ for the LETG, 45\,cm$^{2}$ for the RGS). In addition, unlike
the LETG the RGS is always on (it has a dedicated telescope, whereas the
gratings must be moved into the focal plane on \textit{Chandra}), 
and thus has accumulated many more spectra. These factors lead to many more
detected \ion{O}{7} lines in the RGS, so we only use the LETG data towards
several calibration sources as a check on the wavelength solution (see below).

Our analysis sample includes 37 known \ion{O}{7} absorbers at $z=0$ with
RGS data \citep{nicastro02,fang02,mckernan04,williams05,yao05,fang06,williams06,williams07,yao09,gupta12,fang15}
(Table~\ref{table.velocities}). These include AGNs as well as several X-ray
binaries in the Milky Way's halo and Magellanic Clouds. We tried to include
all sources known to be outside the disk with reported absorption lines, but we
excluded three sources: NGC~3783, PKS~$2005-489$, and
Swift~J$1753.5-0127$. NGC~3783 has an intrinsic oxygen line with a
P~Cygni profile where we cannot disentangle the Galactic line,
PKS~$2005-489$ has a broad line that suggests blending or a
non-Galactic origin, and the line in Swift~J$1753.5-0127$ is only
detected in two of four high $S/N$ exposures. 
We include NGC~5408 X-1 (which is
an X-ray binary, not an AGN)
and NGC~4051, but these systems have redshifts smaller than 1000\,km\,s$^{-1}$ so
the lines may be intrinsic. NGC~4051 and MCG-6-30-15 also have known outflows;
as they have \ion{O}{7} lines attributed to the Galaxy in some prior studies \citep[e.g.,][]{fang15}
we include them here, but we show below that excluding them does not 
strongly change our results.

The data for each target were reprocessed using standard methods in the
\textit{XMM-Newton} Science Analysis Software (SAS v14.0.0) with the
appropriate calibration files. This included excluding hot, cold, and ``cool''
pixels, and data from periods when the background count rate exceeds 3$\sigma$ from the
mean. We applied the (default) empirical correction for the Sun angle of the 
spacecraft and its heliocentric motion \citep{devries15}. We used the
highest precision coordinates available rather than the proposal coordinates,
which improves the accuracy of the wavelength scale. 
For each object, we
merged the first-order RGS1 spectra and response matrices
into a ``stacked'' spectrum. Standard processing resamples
the data from native bins (about 0.011\AA\ at 21.6\AA)
into a user-specified bin size. We binned
the data to 0.02\AA\ (one resolution element is about 0.055\AA). 
This resampling causes small but stochastic changes in the bin assignment for some
events, leading to variation under the same protocols, which we quantify by
processing each object ten times in the same way. 

\subsection{Velocity Measurements}

To measure the Doppler shifts, we fit a model consisting of a power
law and an absorption line to the spectrum in the 21-22\AA\ bandpass
using XSPEC v12.9.0 \citep{arnaud96}. We exclude an instrumental
artifact between 21.75--21.85\AA\ in each
spectrum, and in several spectra there are one or more
bad channels in the bandpass that we also exclude\footnote{see
  http://xmm.esac.esa.int/external/xmm\_user\_support/documentation/uhb/rgsmultipoint.html}.
The parameters of the absorption-line model include the line centroid,
the line width, and the line strength, but we fix the line width at
the instrumental line-profile width because we do not expect detectable
line broadening, an assumption we validated in the brightest sources. 
The best-fit centroid is
converted to a velocity using the best-fit line energy from a stacked
spectrum of Capella as
a reference, corrected for the radial velocity of the star (described below).
We measured the velocity for each of the ten stacked spectra
per object, and we report the mean value with its 1$\sigma$ uncertainty 
in Table~\ref{table.velocities}, including the resampling uncertainty.

\subsection{Wavelength Scale Accuracy}

The systematic uncertainty in the wavelength scale limits the accuracy of
our measurements, and recent improvements in the calibration of the
wavelength scale \citep{devries15} and multiple high $S/N$ spectra for
the objects in our sample are largely what enable this study. Here
we show the accuracy of the wavelength solution for the protocols we adopt
and briefly describe the sources of uncertainty and their magnitudes.	

We created spectra for the active stars Capella and HR~1099 following
the protocols above, then measured the (emission) line centroids for strong,
mostly unblended lines, and
compared them to their laboratory rest wavelengths
(Figure~\ref{figure.capella_wavelength}). We accounted for the radial
velocity of each star \citep[$+29.2$\,km\,s$^{-1}$ for Capella and
  $-15.3$\,km\,s$^{-1}$ for HR~1099;][]{karatas04}. We find no
systematic offset in the 5-30\,\AA\ bandpass or change over time.  The
wavelength offsets in \ion{O}{7} $\lambda$21.602\AA\ for stacked
spectra are $\Delta \lambda = 0.0\pm0.8$\,m\AA\ (Capella) and
$0.7\pm1.6$\,m\AA\ (HR~1099).  These correspond to $v = 0\pm11$ and
$-7\pm22$\,km\,s$^{-1}$. To verify that stacking does not introduce
artificial offsets, we also measured centroids in each individual
exposure (16 for Capella and 14 for HR~1099) and computed the weighted
mean. The offsets are consistent with the stacked spectrum:
$v=-8\pm26$ and $-3\pm28$\,km\,s$^{-1}$ for Capella and HR~1099.

\begin{figure*}
\begin{center}
\hspace{-0.4cm}\includegraphics[width=0.5\textwidth]{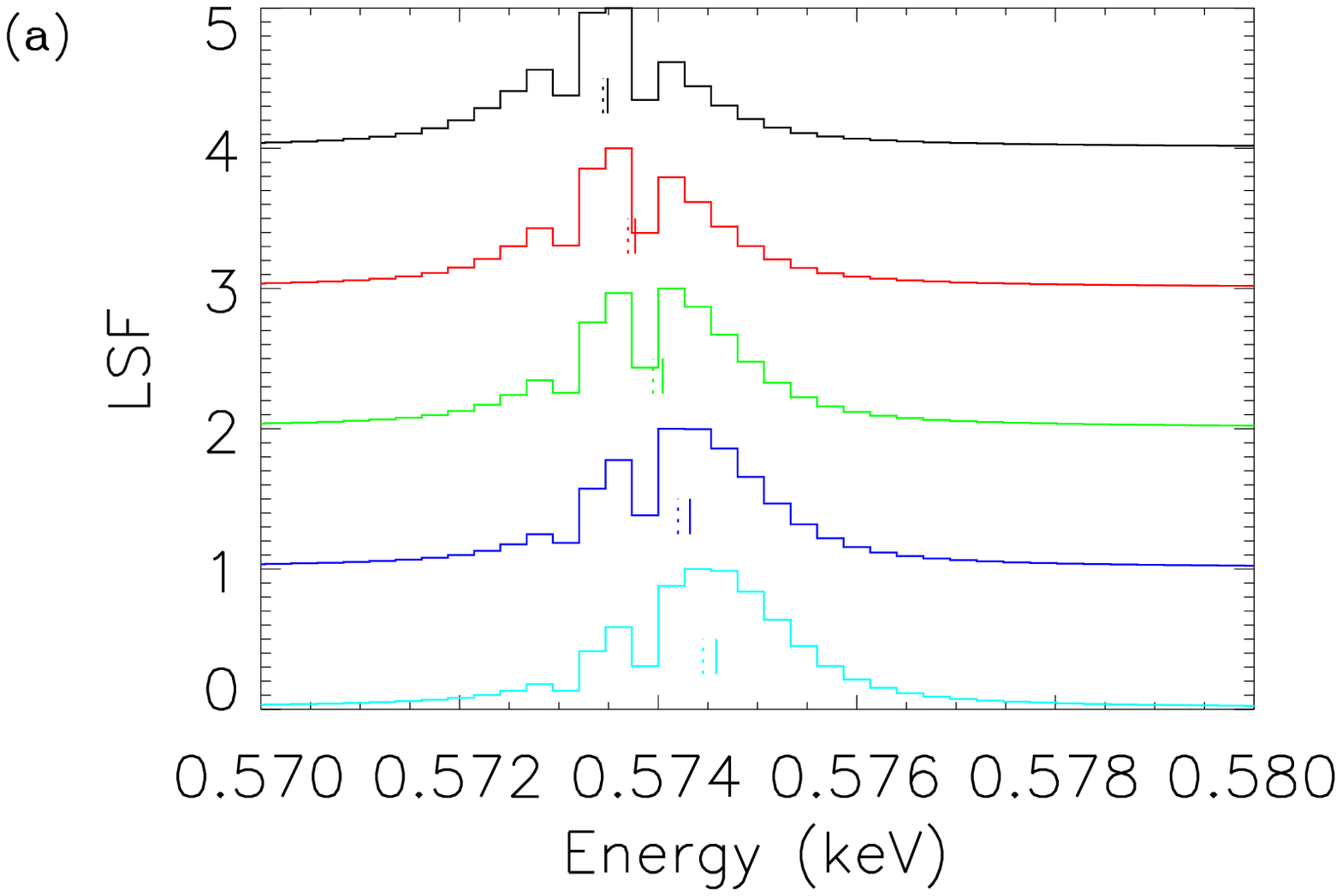}
\hspace{-0.4cm}\includegraphics[width=0.5\textwidth]{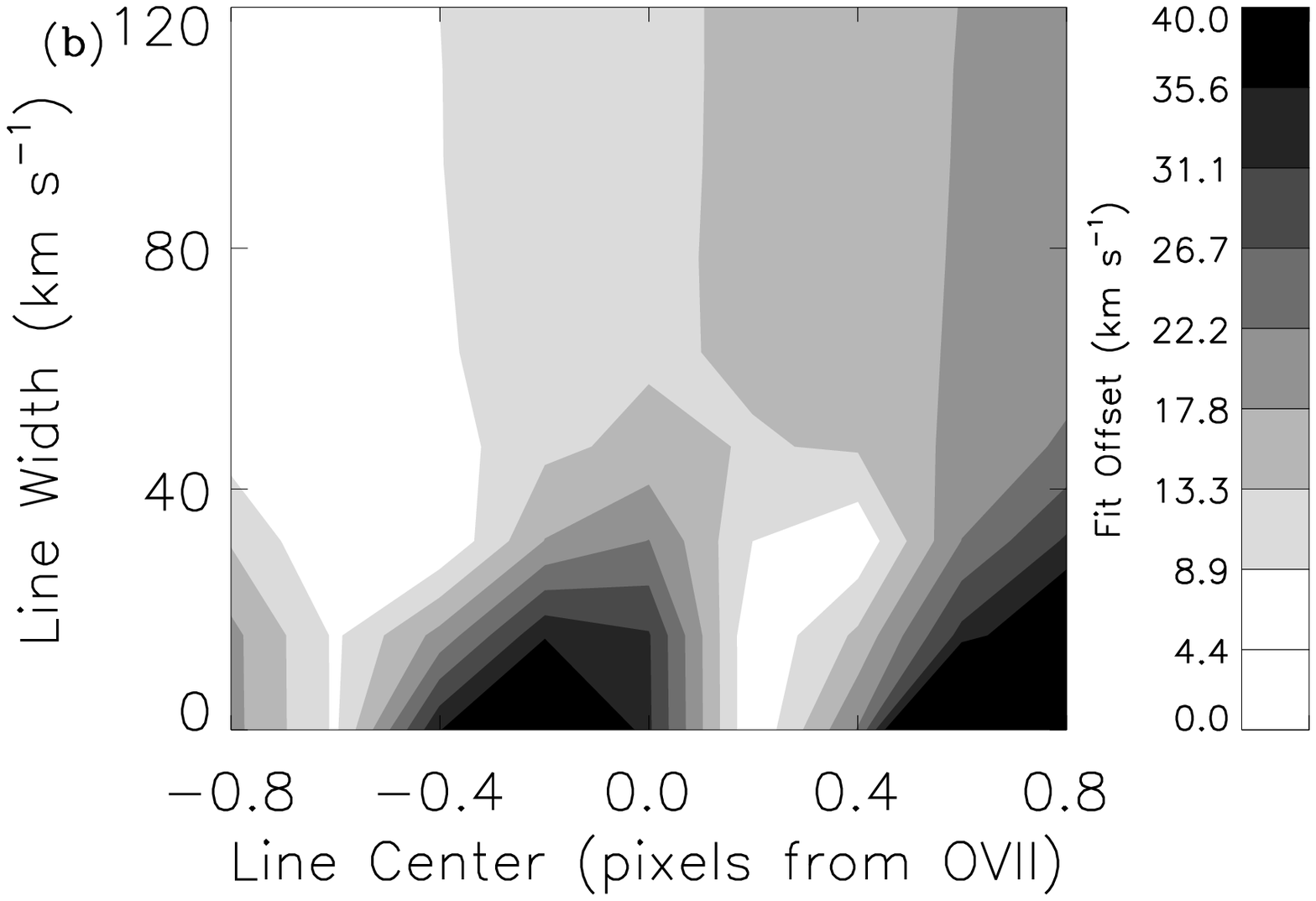}
\caption{Bad columns near the \ion{O}{7} line can cause a
  systematic error in the centroid. (a) The difference between the
  nominal line center (dashed vertical tick) and the measured 
  centroid (solid vertical tick) is a function of the true
  centroid. (b) The magnitude of this offset is plotted as a function
  of intrinsic line width and the nominal line centroid for
  \textit{Xspec} simulations. For lines with an intrinsic width of
  40\,km\,s$^{-1}$ or more, the typical error is about 15\,km\,s$^{-1}$. 
}
\label{figure.lsf_error}
\end{center}
\end{figure*}

We also checked the wavelength solution against the LETG and HETG data for these
stars. We reduced te data using the Chandra Interactive Analysis of
Observations software (CIAO~v4.7), and we found good agreement with
spectra from the reduced data available through the TGCat project
\citep{huenemoerder11}.  We co-added the $\pm$first-order spectra and
stacked all observations. The LETG offsets are $v=1\pm
22$\,km\,s$^{-1}$ (Capella) and $v=10\pm 26$\,km\,s$^{-1}$ (HR~1099),
whereas the HETG Medium Energy Grating offsets are $v =
-3\pm21$\,km\,s$^{-1}$ (Capella) and $v = -22\pm38$\,km\,s$^{-1}$
(HR~1099). These results agree with the RGS and the laboratory
wavelength (Figure~\ref{figure.capella_wavelength}).  We also compared
velocity measurements between the RGS and LETG in the four brightest
quasars
\citep{rasmussen03,williams05,bregman07,rasmussen07,hagihara10,fang15}.
The LETG and RGS centroids agree to within the 1$\sigma$ error bar.

These results show that the wavelength scale is sufficiently accurate
for our measurement, but it is important to note that there is a
substantial systematic scatter that makes \textit{individual}
observations unreliable, and also that there will be systematic
differences between our measurements and those reported for the same
data using an earlier version of SAS or using incorrect source
coordinates.

Different observations of the same object have a systematic scatter of
$\sim$5\,m\AA\ in the wavelength solution around the true mean
value. This corresponds to 70\,km\,s$^{-1}$ at 21.602\AA, and leads to
the standard quoted systematic uncertainty\footnote{see the \textit{XMM-Newton} Users' Handbook at https://heasarc.gsfc.nasa.gov/docs/xmm/uhb/rgs.html}
of 100\,km\,s$^{-1}$. The reason for this scatter is not clear, but
about 3\,m\AA\ could be explained by limits in the pointing accuracy of the
telescope \citep{devries15}. In any case, the scatter is normally
distributed (based on measurements from many exposures of calibration stars such as
Capella), and can thus be strongly mitigated with multiple
observations of the same source. To reduce the scatter to within
20\,km\,s$^{-1}$ requires 15-20 independent spectra (assuming equal
$S/N$ and a $\sigma=5$\,m\AA). The sources in our sample have between
2-60 observations, and at the low end the reported 1$\sigma$
statistical errors (200\,km\,s$^{-1}$ or more) are much larger than
a 70\,km\,s$^{-1}$ systematic error. For the bright quasars with
many observations, we estimate a typical systematic error due to this
scatter of $10-15$\,km\,s$^{-1}$. 

\begin{figure}
\begin{center}
\hspace{-0.4cm}\includegraphics[width=0.5\textwidth]{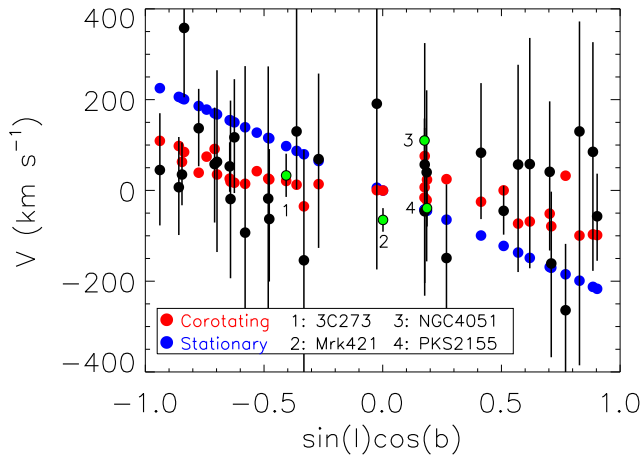}
\caption{RGS measurements of \ion{O}{7} velocity offsets as a function of $\sin(l)\cos(b)$.
Two models are shown for comparison with the data (black points): a stationary
model (blue points) and a corotating model (red points). Error bars are 1$\sigma$ (standard deviation). 
The four objects with the
smallest error bars are labeled 1-4. The velocities
of the lines towards Mrk~421 and NGC~4051 (labeled `2' and `3') suggest
some intrinsic scatter.
}
\label{figure.velocities}
\end{center}
\end{figure}

In addition to the scatter, there are systematic offsets produced by
the Sun angle of the telescope and its projected heliocentric motion
(possibly an inaccuracy in the star-tracker) that were measured and
corrected by \citet{devries15}; we refer the reader to their paper for a
  detailed description. The earliest version of SAS that
conained these corrections was v13.0.0, and it was not enabled by
default until v14.0.0; all prior absorption-line halo studies using
the RGS used an earlier version or did not mention the correction. 
We measured line centroids with and without these corrections, and found
centroid shifts in our sample with magnitudes between $0-100$\,km\,s$^{-1}$ (higher for weak
lines). To show the effect of not
including them, we show the measured centroids without these
corrections in Table~\ref{table.velocities}. Finally, if we used the default
(proposal) coordinates instead of the SIMBAD coordinates, we measured
offsets of $\pm10$\,km\,s$^{-1}$. 

Thus, we would expect our measured centroids to be correlated with
prior results but perhaps significantly different relative to the
statistical errors. For example, there is a systematic offset of about
60\,km\,s$^{-1}$ between our measurements and \citet{fang15} for the
several systems where it can be measured. This offset cannot be
entirely explained by the Sun angle and heliocentric motion
corrections, but it is consistent with shifts seen relative to prior
versions of SAS \citep{devries15}. They do not report the line
centroids for active stars, as they only need to measure the centroids
to sufficient accuracy to identify halo absorbers. Thus, we do not
regard the apparent inconsistency as reflecting an inherent
uncertainty in line centroid measurements.

\begin{deluxetable*}{lcccccccccc}
\tablenum{1}
\tabletypesize{\scriptsize}
\tablecaption{Oxygen Absorption Line Sample}
\tablewidth{0pt}
\tablehead{
\colhead{Name} & \colhead{Gal Long.} & \colhead{Gal Lat.} & \colhead{$z$} &
\colhead{$\sin(l)\cos(b)$} & \colhead{$v_0$} & \colhead{$v_0 -
  v_{0,\text{nc}}$} & \colhead{Corotating} & \colhead{Stationary} & \colhead{Eq. Width} &
\colhead{Cont. $S/N$}  \\
\colhead{} & \colhead{(deg)} & \colhead{(deg)} & \colhead{} & 
\colhead{} & \colhead{(km s$^{-1}$)} & \colhead{(km s$^{-1}$)} & 
\colhead{(km s$^{-1}$)} & \colhead{(km s$^{-1}$)} & \colhead{(m\AA)} 
}
\startdata
Mkn 421			& 179.832	& 65.031	& 0.0300 & $0.0013$	 & $-65_{-21}^{+21}$ 	& $-29$	 &	$0$		& $0$    & 13.1$\pm$0.5 & 232 	\\
3C 273			& 289.951	& 64.360	& 0.1580 & $-0.4069$ & $33_{-45}^{+45}$ 	& $-27$	 &	$21$	& $98$   & 24.6$\pm$1.6 & 55 	\\
PKS 2155-304	& 17.730	& $-52.245$	& 0.1160 & $0.1865$	 & $-39_{-38}^{+38}$ 	& $-36$	 &	$24$	& $-45$  & 15.4$\pm$0.9 & 95 	\\
Mkn 509			& 35.971	& $-29.855$	& 0.0340 & $0.5094$	 & $-45_{-50}^{+53}$ 	& $-41$	 &	$0$		& $-122$ & 30.2$\pm$2.4 & 48 	\\
NGC 4051		& 148.883	& 70.085	& 0.0023 & $0.1761$	 & $110_{-41}^{+46}$ 	& $8$	 &	$-16$	& $-42$  & 40.4$\pm$3.6 & 33 	\\
MCG -06-30-15	& 313.292	& 27.680	& 0.0077 & $-0.6448$ & $53_{-54}^{+51}$ 	& $-1$	 &	$26$	& $154$  & 35.1$\pm$3.2 & 38 	\\
Ark 564			& 92.138	& $-25.337$	& 0.0247 & $0.9032$	 & $-57_{-97}^{+93}$ 	& $21$	 &	$-98$	& $-217$ & 12.3$\pm$1.9 & 50 	\\
ESO 141-55		& 338.183	& $-26.711$	& 0.0371 & $-0.3323$ & $-154_{-264}^{+242}$ & $102$	 &	$-35$	& $80$   & 21.2$\pm$5.1 & 18 	\\
H1426+428		& 77.487	& 64.899	& 0.1290 & $0.4142$	 & $83_{-145}^{+153}$ 	& $-7$	 &	$-25$	& $-99$  & 16.0$\pm$3.9 & 25 	\\
1H 0707-495		& 260.169	& $-17.672$	& 0.0405 & $-0.9388$ & $45_{-121}^{+124}$ 	& $-21$	 &	$109$	& $225$  & 24.3$\pm$6.1 & 22 	\\
PKS 0558-504	& 257.962	& $-28.569$	& 0.1370 & $-0.8589$ & $7_{-104}^{+110}$ 	& $8$	 &	$98$	& $206$  & 16.6$\pm$4.2 & 32 	\\
NGC 4593		& 297.483	& 57.403	& 0.0090 & $-0.4781$ & $-63_{-137}^{+153}$ 	& $11$	 &	$24$	& $115$  & 27.1$\pm$7.1 & 13  	\\
Mrk 335			& 108.763	& $-41.424$	& 0.0258 & $0.7100$	 & $-161_{-206}^{+158}$ & $-40$	 &	$-79$	& $-170$ & 19.3$\pm$5.3 & 22 	\\
PG 1211+143		& 267.552	& 74.315	& 0.0809 & $-0.2702$ & $69_{-195}^{+188}$ 	& $-57$	 &	$14$	& $65$   & 40.3$\pm$11. & 7  	\\
PG 1244+026		& 300.041	& 65.214	& 0.0482 & $-0.3631$ & $130_{-127}^{+286}$ 	& $-188$ &	$12$	& $87$   & 40.8$\pm$11. & 10  	\\
Mkn 501			& 64.600	& 38.860	& 0.0337 & $0.7034$	 & $41_{-173}^{+155}$ 	& $-68$	 &	$-51$	& $-169$ & 25.2$\pm$10. & 22 	\\
1ES 1028+511	& 161.439	& 54.439	& 0.3600 & $0.1853$	 & $40_{-196}^{+180}$ 	& $81$	 &	$-21$	& $-44$  & 36.4$\pm$11. & 12 	\\
ESO 198-24		& 271.639	& $-57.948$	& 0.0455 & $-0.5305$ & $-652_{-219}^{+231}$	& $-393$ &	$42$	& $127$  & 62.2$\pm$19. & 7  	\\
NGC 2617		& 229.300	& 20.939	& 0.0142 & $-0.7079$ & $58_{-128}^{+124}$ 	& $-95$	 &	$91$	& $170$  & 30.1$\pm$9.5 & 10  	\\
PG 1553+113		& 21.909	& 43.964	& 0.3600 & $0.2686$	 & $-149_{-175}^{+162}$ & $45$	 &	$25$	& $-64$  & 27.6$\pm$8.7 & 13  	\\
H1101-232		& 273.190	& $33.079$	& 0.1860 & $-0.8367$ & $358_{-152}^{+141}$	& $2$	 &	$85$	& $201$  & 45.5$\pm$17. & 8  	\\
3C 390.3		& 111.438	& 27.074	& 0.0561 & $0.8289$	 & $130_{-515}^{+242}$ 	& $-21$	 &	$-99$	& $-199$ & 25.0$\pm$10. & 10  	\\
1H 0419-577		& 266.963	& $-42.006$	& 0.1040 & $-0.7421$ & $783_{-255}^{+273}$	& $17$	 &   $74$	& $178$  & 39.9$\pm$16. & 6  	\\ 
Fairall 9		& 295.073	& $-57.826$	& 0.0470 & $-0.4825$ & $-18_{-333}^{+291}$ 	& $-226$ &	$26$	& $116$  & 37.2$\pm$15. & 6  	\\
IRAS 13224-3809	& 310.189	& 23.979	& 0.0660 & $-0.6982$ & $63_{-198}^{+201}$	& $31$	 &	$35$	& $167$  & 45.1$\pm$19. & 7  	\\
NGC 5408 X-1	& 317.149	& 19.496	& 0.0017 & $-0.6414$ & $-19_{-174}^{+216}$ 	& $-98$	 &	$19$	& $154$  & 30.4$\pm$14. & 7  	\\
PDS 456			& 10.392	& 11.164	& 0.1840 & $0.1770$	 & $57_{-261}^{+267}$ 	& $47$	 &	$76$	& $-42$  & 60.0$\pm$29. & 7  	\\
Mrk 279			& 115.042	& 46.865	& 0.0305 & $0.6195$	 & $58_{-229}^{+278}$ 	& $-29$	 &	$-68$	& $-149$ & 14.2$\pm$7.0 & 15  	\\
E1821+643		& 94.003	& 27.417	& 0.2970 & $0.8855$	 & $85_{-262}^{+241}$ 	& $-28$	 &	$-97$	& $-212$ & 35.7$\pm$18. & 7  	\\
IC 4329A		& 317.496	& 30.920	& 0.0160 & $-0.5799$ & $-93_{-229}^{+367}$ 	& $-73$	 &	$15$	& $139$  & 38.3$\pm$20. & 7  	\\
PG 0804+761		& 138.279	& 31.033	& 0.1000 & $0.5704$	 & $57_{-236}^{+219}$ 	& $98$	 &	$-73$	& $-137$ & 28.5$\pm$15. & 6  	\\
NGC 5548		& 31.960	& 70.496	& 0.0172 & $0.1768$	 & $-46_{-185}^{+183}$ 	& $-51$	 &	$7$		& $-42$  & 18.0$\pm$10. & 13  	\\
Mrk 766			& 190.681	& 82.270	& 0.0129 & $-0.0249$ & $191_{-365}^{+240}$ 	& $93$	 &	$0$		& $6$    & 6.0$\pm$4.2  & 27 	\\
\cutinhead{Local Group and Halo Sources}
LMC X-3			& 273.576	& $-32.082$	& -      & $-0.8457$ & $35_{-66}^{+69}$		& $7$	 &	$63$	& $203$  & 23.1$\pm$3.0 & 36 	\\
4U 1957+11		& 51.308	& $-9.330$	& -      & $0.7702$	 & $-264_{-152}^{+145}$ & $-124$ &	$32$	& $-185$ & 20.6$\pm$4.1 & 22 	\\
MAXI J0556-032	& 238.939	& $-25.183$	& -	 	 & $-0.7751$ & $137_{-82}^{+85}$ 	& $84$	 &	$39$	& $186$  & 16.0$\pm$3.3 & 33 	\\
SMC X-1			& 300.414	& $-43.560$	& -      & $-0.6251$ & $117_{-122}^{+127}$ 	& $-3$	 &	$17$	& $150$  & 21.0$\pm$4.9 & 13 	\\
\enddata
\tablecomments{\label{table.velocities} 
The quantity $v_{0} - v_{0,\text{nc}}$ is the difference between measurements with and without the 
heliocentric/sun-angle corrections. The corotating and stationary columns refer to model velocities. 
The equivalent width is for a Lorentzian line of fixed width (0.028\AA), and 
the continuum $S/N$ is per resolution element ($\sim$0.055\AA).
}
\end{deluxetable*}

\subsection{Other Systematics}

In addition to the wavelength grid, there are a few sources of
systematic uncertainty and some fitting choices that affect our final
results but where we believe there is a correct choice. We briefly
describe these here.

\paragraph{\textit{Cool Pixels}} There are several ``cool'' pixels in
the vicinity of 21.6\AA\ that have a lower than expected signal (by about 20\%). By
default, these pixels are included in the spectrum. We exclude them
because they can affect weak absorption features. The typical velocity
shift between keeping and excluding them is $\Delta v = 10$\,km\,s$^{-1}$
in bright sources.

\paragraph{\textit{Binning and Resampling}} The native RGS binning is
about 11\,m\AA\ at 21.6\AA, but the default processing resamples the events onto a
user-specified grid, with a default value of 10\,m\AA. This resampling
is probabilistic with a random element (normally distributed), which
means that running the same protocol on a given data set multiple
times will result in slightly different spectra. The magnitude of the
velocity shift is strongly dependent on the line strength and the
continuum $S/N$, so we reprocessed each data set ten times per
reduction protocol set. We then added the standard deviation in the
measured velocities in quadrature to the statistical error, since it
behaves in essentially the same way. The bin sizes also affect the
measured velocities, with a mean line centroid shift of $\Delta v =
7$\,km\,s$^{-1}$ between bins of 10\,m\AA\ and 20\,m\AA\ in bright
systems (we use the latter). 

\paragraph{\textit{Stacking}} Temporal changes in the instrumental
response or changes in the spectral shape of the source can bias the
results of stacked spectra. On the other hand, jointly fitting several
low-$S/N$ spectra with the continuum shape as a free parameter 
leads to poorer constraints on the velocity of a line based on
spectral bins far from the absorption line. We stack the spectra to
improve the continuum $S/N$, but to determine if this provides
systematic bias we measured the line shift between joint fits and
stacked spectra in our brightest sources and the calibration
sources. We find a typical $\Delta v = 5$\,km\,s$^{-1}$ because the
instrumental response in these regions does not appear to have
temporal changes.

\begin{figure*}
\begin{center}
\includegraphics[width=0.9\textwidth]{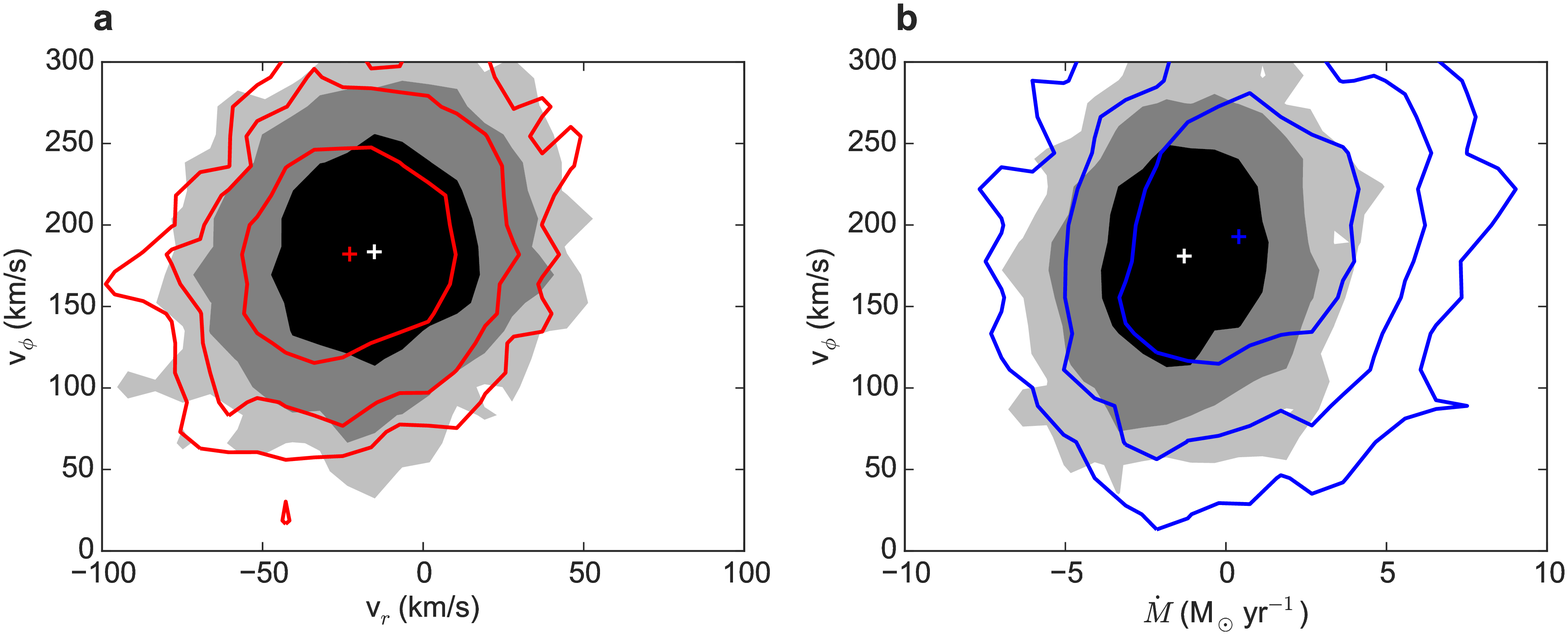}
\includegraphics[width=0.9\textwidth]{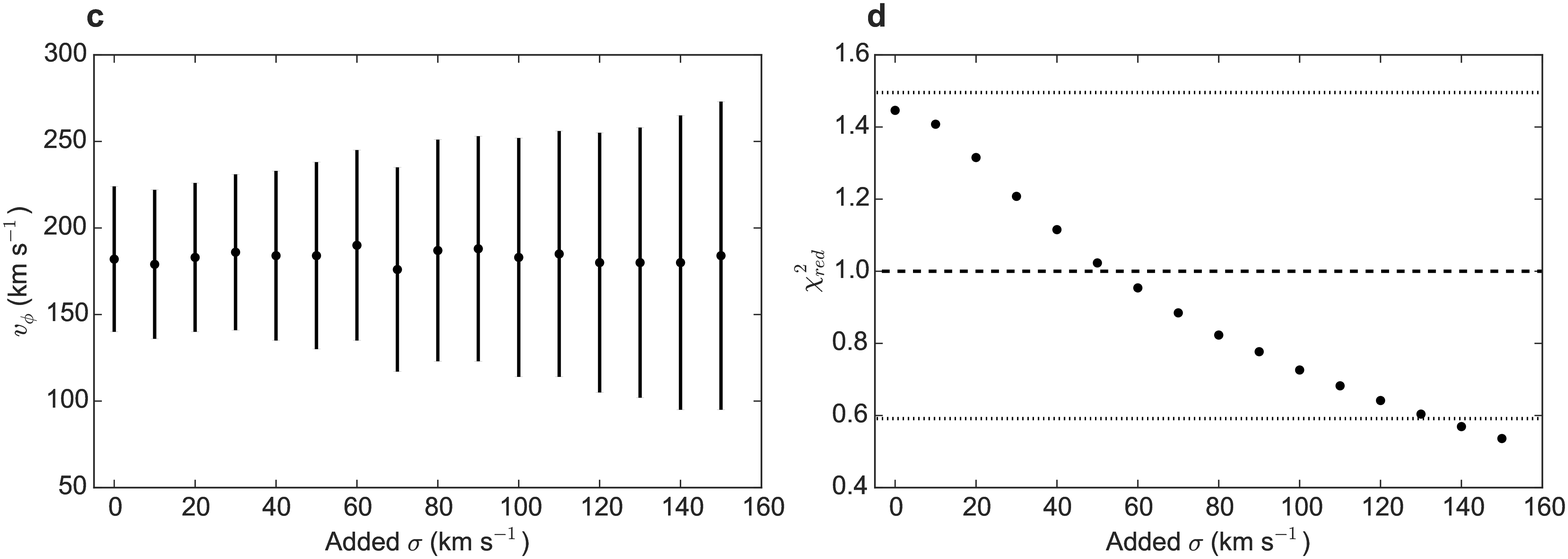}
\caption{
(a) Contours show the 1, 2, and 3$\sigma$ deviations from the best-fit 
parameters for the whole sample (shaded) and lines with measurement errors
less than 250\,km\,s$^{-1}$ (lines). (b) The radial velocity can be converted 
to an accretion or outflow rate, which is consistent with zero. 
(c) When the possibility of intrinsic
scatter is included in the model, $v_{\phi}$ becomes less certain, but even 
with 150\,km\,s$^{-1}$ of added scatter, some rotation is necessary. (d)
The added scatter value that brings the reduced $\chi^2$ closest to 1.0 is near
50\,km\,s$^{-1}$, with values above 140\,km\,s$^{-1}$ ruled out with 90\% confidence
(dotted lines).
}
\label{figure.results}
\end{center}
\end{figure*}

\paragraph{\textit{Line Profile and Fixed Line Width}} For lines with Doppler $b <
200$\,km\,s$^{-1}$, we do not expect to measure reliable line widths
\citep[however, see the Doppler $b$ measurements in some of the same
  lines in][]{fang15}. Since the RGS line-spread function is best
described as a Lorentzian near the core, we fit our spectra using a
Lorentzian profile with the width fixed at the instrumental line
width. If we use a Gaussian line profile instead but keep the width
fixed, the line shift is consistent with zero in most cases but can be
up to $\Delta v = 30$\,km\,s$^{-1}$ in weak lines (statistical errors
greater than 200\,km\,s$^{-1}$). When the line width is a free
parameter, we find shifts in the Lorentzian centroids of $\Delta v =
0-15$\,km\,s$^{-1}$ (the shift magnitude is negatively correlated with
$S/N$) and $\Delta v = 20-50$\,km\,s$^{-1}$ in the Gaussian
case. However, for the Gaussian lines the best-fit line widths tend to
be $1.5-2$~times the instrumental resolution, and the statistical
error also increases. These fits are usually not significantly better
than with a fixed line width, so in our view a non-zero line width is
not required by the data, and these line widths reflect some small curvature
in the continuum. 

\paragraph{\textit{Bandpass}} The fitting bandpass is important
because the continuum model needs to fit well even if the result is
unphysical. Typically, one fits the local continuum, but in the
literature for the \ion{O}{7} line this can vary from a fitting
interval less than 1\,\AA\ wide to about 5\AA\ wide. Our choice of
fitting bandpass (21-22\AA) is motivated by strong instrumental
features below 21\AA and above 22\AA, but if we ignore these features
and expand our bandpass by $\pm$1\AA, the typical velocity shift is
$\Delta v = 3$\,km\,s$^{-1}$. 

\paragraph{\textit{Bad Columns near 21.6\AA}} There is asymmetry in the
line-spread function (LSF), which is the instrumental response to a
$\delta$-function, near the \ion{O}{7} line. This is not the same as
the well known asymmetry in the LSF at the 1\% level in the line wings (which is
not important), but rather due to bad columns that are not included in the
cool pixel list
(Figure~\ref{figure.lsf_error}). For an arbitrarily strong
$\delta$-function line, the offset in the measured centroid from this
defect can range from $\Delta v = 0-100$\,km\,s$^{-1}$ depending on
where the true line centroid is. However, if the line is unresolved
but has some physical width (so that incident photons would be
dispersed over multiple bins even before the LSF spreads them out),
the error is strongly mitigated. We used XSPEC simulations to
determine the error as a function of Doppler $b$ parameter, and for
$b>40$\,km\,s$^{-1}$ (about the thermal width even without
turbulence), the typical error is reduced to 15\,km\,s$^{-1}$
(Figure~\ref{figure.lsf_error}).  We expect the Galactic halo lines to
be in this regime. Alternatively, one can ignore the columns, but
because of their proximity to the region of interest, this will also
bias the results.

Overall, the systematic error should allow measurements to better than
30\,km\,s$^{-1}$ accuracy with sufficient $S/N$. For the high $S/N$
lines we estimate a typical systematic error of 15-20\,km\,s$^{-1}$,
and for lines weaker than about 4$\sigma$, the systematic error is
dominated by the statistical error.

\section{Halo Model}

We used a simple halo model for comparison to the data. We adopted an
extended density profile \citep{miller15} in which
\begin{equation}
n(r) = n_0(1+(r/r_c)^2)^{-3\beta/2} \text{ cm$^{-3}$}
\label{equation.beta_model}
\end{equation}
where $n_0 r_c^{3\beta} =
1.35\pm0.24$\,cm$^{-3}$\,kpc$^{3\beta}$ and $\beta = 0.50\pm0.03$,
computed on a grid with 0.05\,kpc cells.  This model describes a large,
all-sky sample of 649 \ion{O}{7} and O{\sc viii} emission lines
\citep{henley10,henley12} well. We then imposed global bulk radial
($v_r$) or azimuthal ($v_{\phi}$) velocities, assuming that these are
constant with radius. These are the free parameters in the model.  We
likewise assume a constant metallicity ($Z=0.3 Z_{\odot}$)
and a Doppler $b$ parameter of 85\,km\,s$^{-1}$ due to random
turbulent motion in each cell, based on hydrodynamic simulations
\citep{cen12}. We also account for the possibility of intrinsic
scatter (resulting from hydrodynamic flows) about any model by adding
a velocity dispersion to the velocities when comparing to the models.
This dispersion is not the same as line \textit{broadening}, but refers to
the typical value for a distribution of centroid shifts.

\begin{figure*}
\begin{center}
\hspace{-0.25cm}\includegraphics[width=0.45\textwidth]{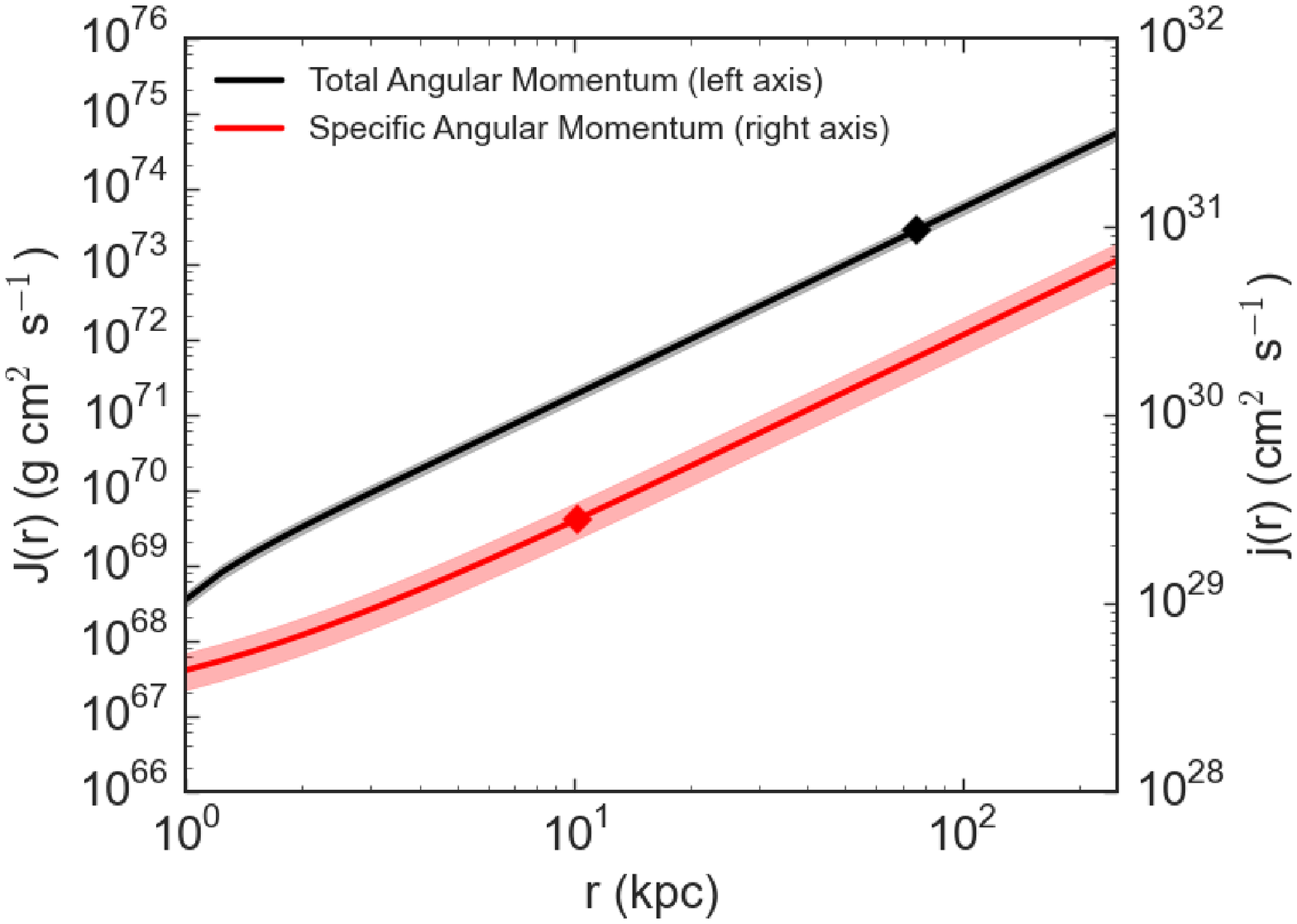}
\includegraphics[width=0.45\textwidth]{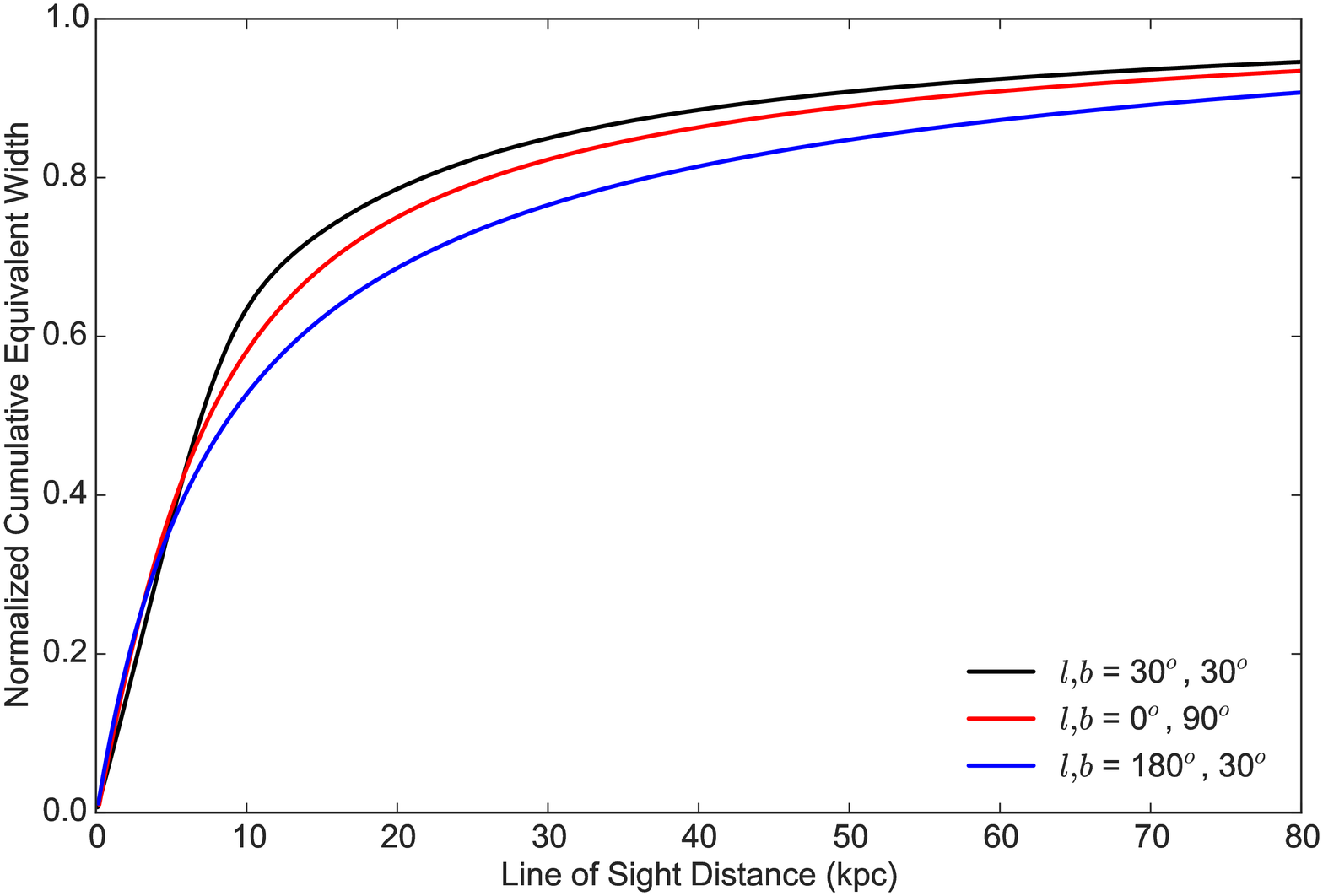}
\caption{
(a) The black (red) line shows the total (specific) angular momentum of
the hot gas as a function of radius in our density model. Both assume
$v_{\phi} = 183\pm41$\,km\,s$^{-1}$. The diamonds show the values for the sum of stellar and H{\sc i}
values in the disk of the Milky Way. (b) The cumulative equivalent width for
three sightlines shows that, while different sightlines are sensitive to gas
at different distances, in all cases more than 70\% of the equivalent width
comes from beyond 5\,kpc from the Sun.
}
\label{figure.momentum}
\end{center}
\end{figure*}

We obtain model velocities for each object by integrating from the
position of the Sun outward along each line of sight, computing the
line-of-sight velocity component and broadening for each cell and
summing the resultant Voigt profiles \citet[a more detailed description is
given in][]{miller16}. We convolve the result with the
LSF, compute the centroid, and add the solar reflex motion for
comparison to the observations. 
In other words, the velocities are compared in the frame of the Local Standard
of Rest.
For this model, at least 50\% of the
total absorption comes from beyond 7\,kpc from the Sun, and 90\% from 
within 50\,kpc. 
In the simplest case of a stationary halo, the Galactic
rotation of the Local Standard of Rest \citep{reid14} ($v_{\text{LSR}}
= 240\pm 8$\,km\,s$^{-1}$) is reflected in the measured Doppler
shifts: $v_{\text{obs}} = -240\sin(l)\cos(b)$, where $l$ is the
Galactic longitude and $b$ the Galactic latitude. The product
$\sin(l)\cos(b) = \pm1$ corresponds to the Sun moving directly towards
or away from that direction, resulting in a Doppler shift of
$\mp240$\,km\,s$^{-1}$. Another simple case is a corotating halo
($v_{\phi} = v_{\text{LSR}}$), in which case the Doppler shifts will
be closer to zero.  Figure~\ref{figure.velocities} shows the measured velocities with the
stationary and corotating models. 

\section{Results}

The best-fit $v_r$ and $v_{\phi}$ values were obtained from a
Markov-chain Monte Carlo (MCMC)
approach, using the $\chi^2$ statistic
as a goodness-of-fit parameter. For zero dispersion, the best model has a prograde rotation
velocity of $v_{\phi} = 183\pm41$\,km\,s$^{-1}$ and a global inflow of
$v_r = -15\pm20$\,km\,s$^{-1}$, corresponding to a net accretion rate of
$\dot{M} = 1\pm2 M_{\odot}$\,yr$^{-1}$
 (Figure~\ref{figure.results}). This is a formally
acceptable fit, whereas the stationary halo is rejected with 99.95\%
probability and the corotating halo is marginally rejected with 95\%
probability. The velocity dispersion for which the reduced $\chi^2$ is
closest to 1.0 is about 50\,km\,s$^{-1}$ (Figure~\ref{figure.results});
substantially more than this produces a $\chi^2$ that is too small for
the observed line centroids. The
apparent inflow is not statistically significant, and the suggestion
of inflow primarily results from Mrk~421 (Figure~\ref{figure.velocities}), which has a small
uncertainty and is near $\sin(l)\cos(b) = 0$.  Taking these results at
face value, the large $v_{\phi}$ and extent of the halo imply that the
total angular momentum of the hot gas within 50\,kpc is comparable to
that in the stars and gas in the disk of the Galaxy \citep{mo10}
(Figure~\ref{figure.momentum}).
The spread in recent measurements of $v_{\text{LSR}} = 200-250$\,km\,s$^{-1}$ 
\citep{brunthaler11,bovy09,reid14} leads to $v_{\phi} = 130-180$\,km\,s$^{-1}$, which 
does not change the picture of a lagging halo with prograde rotation. 

\begin{figure*}
\begin{center}
\hspace{1.5cm}\includegraphics[width=0.75\textwidth]{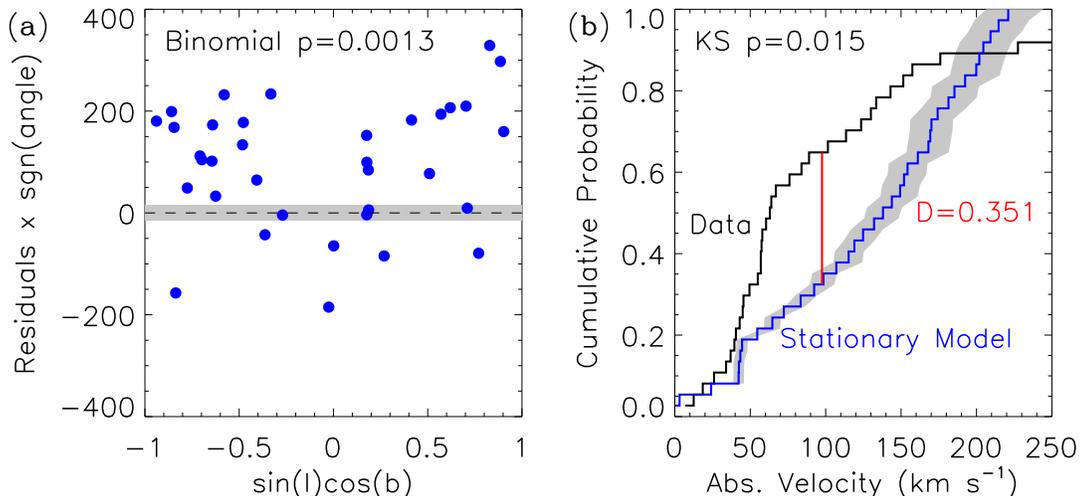}
\caption{\textbf{(a).} The residual distribution around the stationary model shows that too many
sight lines have velocities less than the model predictions. At each angle, a
positive residual indicates that the model has a higher velocity
magnitude than the measured offset.
\textbf{(b).} 
The cumulative distributions of the model and measured velocity
magnitudes (right panel) also rule out a stationary model by the KS
test at the 98\% level. The shaded grey regions in each panel represent
$\pm20$\,km\,s$^{-1}$ from $v_{\text{LSR}} = 240$\,km\,s$^{-1}$.
}
\label{figure.nonpar}
\end{center}
\end{figure*}

In addition to parameter fitting, we used nonparametric statistical tests
to test the hypothesis that some rotation is necessary. We tested the stationary
and corotating models using the sign test and the Kolmogorov-Smirnov (KS)
test, which compare distributions and are less sensitive to scatter. 
The sign test asks whether the medians of the
measured and model velocities are consistent with each other (assuming
a binomial distribution and a 50\% probability that the model velocity
exceeds the measured velocity). The left panel of Figure~\ref{figure.nonpar} shows the residuals
from subtracting the data from the stationary model values, and the strong
asymmetry rules out the stationary model with 99.87\% probability. The
KS test compares the cumulative distributions of the measured and model 
velocities, and this test rules out the stationary halo with 98.5\%
probability. The corotating model is acceptable in the sign test (43\% rejection
probability) and not in the KS test (99.8\% probability). The rejection of the
stationary halo is model independent.
If we exclude weak lines with $\sigma_{\text{stat}} \ge 250$\,km\,s$^{-1}$,
a stationary model is still ruled out at more than 99.4\% probability. If we
exclude ambiguous lines (such as those towards NGC~5408 X-1, NGC~4051,
and MCG-6-30-15), a stationary model is ruled out at about 98\%
probability in the KS test and 99.6\% in the sign test. Excluding NGC~4051
(with its small error bars) increases the range of acceptable 
$v_{\phi}$ in the parametric fits, but the best-fit $v_r$ does not change
much as the error bars for Mrk~421 are even smaller.

We investigated the sensitivity of the best-fit parameters and the
implied angular momentum to the model assumptions. First,
the assumption of constant velocity must break down at some radius. An
effort to measure the Galactic rotation curve to 200\,kpc using disk
and non-disk objects found \citep{bhattacharjee14} that $v_{\phi}$ is
flat to 80\,kpc, while a measurement from disk stars in the Sloan
Digital Sky Survey found \citep{xue08} that $v_{\phi}$ in the disk
declines by 15\% from $v_{\text{LSR}}$ at 10-20\,kpc and remains
constant from 20-55\,kpc (the maximum probed). Thus, the assumption of
a constant $v_{r}$ or $v_{\phi}$ within $r<80$\,kpc appears to be
reasonable, and more than 80\% of the cumulative equivalent width
comes from within this region (Figure~\ref{figure.momentum}). When we
considered a model with a constant $v_{\phi}$ within 50\,kpc and
$v_{\phi}=0$\,km\,s$^{-1}$ beyond, the best-fit $v_{\phi}$ is nearly
identical to the reported value. For a 10\,kpc cutoff, the best-fit
$v_{\phi}$ increases. 

Second, there may be a metallicity gradient in
the hot gas, in which case the cumulative equivalent width will be
even more dominated by nearby gas. This would impact the inferred mass
and total angular momentum of the halo gas. Constraints from O{\sc
  vii} and O{\sc viii} and pulsar dispersion measures towards the
Large Magellanic Cloud are consistent \citep{miller15} with a gradient
of $Z(r) \propto r^{-0.2}$ with $Z=0.39 Z_{\odot}$ at 1\,kpc and
$Z=0.26 Z_{\odot}$ at 10\,kpc. The uncertainty is large, but the
gradient is shallow enough that gas beyond a few kpc contributes 50\%
or more of the equivalent width. 

Finally, even a single-component
hot halo is probably not isothermal and may also have flows arising
from satellite galaxy motions or gaseous inflows and outflows. Since
the combined emission- and absorption-line data favor an extended halo
and its temperature is near the virial temperature, the gas is
probably volume filling.  Hydrodynamic flows in such a halo will
primarily induce scatter of the type described above, which does not
strongly affect our $v_{\phi}$ measurement (Figure~\ref{figure.results}).
Thus, the measured halo rotation probably extends to tens of kpc and
possibly to 100\,kpc, and this gas will have a substantial angular
momentum. However, the angular momentum values (Figure~\ref{figure.momentum})
are strongly model dependent, especially due to the assumption of constant
$v_{\phi}(r)$.

\section{Discussion}

\subsection{Halo and Disk Models}

The interpretation of the measured Doppler shifts as a rotation signature
depends on the validity of our single-component halo model. This 
depends on the following assumptions:
(1) A volume-filling spherical halo is an
approximately accurate representation of most of the hot gas around the
Galaxy; (2) there is no strong local absorber that we have ignored; (3) 
The spherical halo has bulk global motion; (4) the assumption of $dv_{\phi}/dR=0$ 
is reasonable within about 50\,kpc; and (5) the Doppler shift measurements are accurate. 
The fourth and fifth assumptions were addressed above, so here we focus on the
first three.

Two basic models are suggested in the literature for the structure of the hot
halo: an extended, spherical distribution or an exponential disk with a scale
height of a few kpc. When using individual sight lines or small
samples \citep{yao09b,hagihara10}, the observed 
\ion{O}{7} and \ion{O}{8} absorption and emission line strengths are consistent
with an exponential disk model with a scale height of a few kpc. They are
also consistent with a spherical model. However, 
larger samples of absorption lines \citep[$\sim$40 sight lines;][]{bregman07,gupta12,fang15} and the all-sky 
emission-line intensities \citep[$\approx$1,000 sight lines][]{henley12,miller13,miller15}
favor a spherical model. Similar analyses on independent observables that also
probe the hot gas, such as the pulsar dispersion measure towards the
Large Magellanic Cloud and the ram-pressure stripping of Milky Way
satellites favor an extended halo \citep{fang13,salem15}. 
Finally, a recent analysis
of both Galactic and extragalactic sightlines for $L_*$ galaxies finds that the
\ion{O}{7} traces hot gas \citep{faerman16}.
Thus, we adopted the spherical density profile of \citet{miller15}. 

However, these analyses are based on single-component models, 
and from basic galaxy models we expect at
least two X-ray absorbing components: infalling gas that is shock heated to the virial
temperature ($T \approx 2\times 10^6$\,K for
the Milky Way) and forms an extended halo \citep[e.g.,][]{white91}, and 
supernova-driven outflows from the disk \citep[e.g.,][]{shapiro76,hill12}. 
In the latter case, we expect an exponential disk of hot gas with a scale
height set (for a given Galactocentric radius) by the temperature at the 
midplane. It is worth noting that for $T = 2\times 10^6$\,K, the scale height
at the Solar circle is larger than the Galactocentric distance, so the
distribution of outflowing gas could also be spherical, if not very extended;
the true shape depends on how widespread the outflows are in the disk, the
actual midplane temperature, and the amount of radiative cooling. Within the
galaxy disk itself, supernovae contribute to the hot interstellar medium, much
of which is confined within the disk. Since X-ray absorption covers a wider temperature
range than emission, we are also weakly sensitive to cooler gas (described in 
the following subsection).

These components may be kinematically distinct, as is seen
in the warm gas by \citet{nicastro03,savage03}, but at the relatively low
resolution of the RGS, we expect them to be blended. This complicates the
interpretation of the measured centroids, which are the weighted average of the
offsets for each component along that line of sight. At a qualitative level,
we expect the gas confined in the galaxy disk to rotate with it (although
depending on the distances to the absorbers, there may not be any
rotation signature) and the gas in
supernova-driven outflows to rotate in the same direction but lagging the disk
as it reaches larger heights or radii. 

To constrain the column and thus the influence on the line centroids from a
disk component, we extended the analysis of
\citet{miller15} to fit a disk$+$halo model to the same data set they used:
648 \ion{O}{8} emission lines from \citet{henley12}, which are filtered for
contamination from solar-wind charge exchange and ignore most of the Galactic
plane. We refer the reader to \citet{miller15}
for a more detailed explanation of the modeling procedure, but we
summarize our model components here. 

\begin{figure*}
\begin{center}
\hspace{-0.4cm}\includegraphics[width=0.9\textwidth]{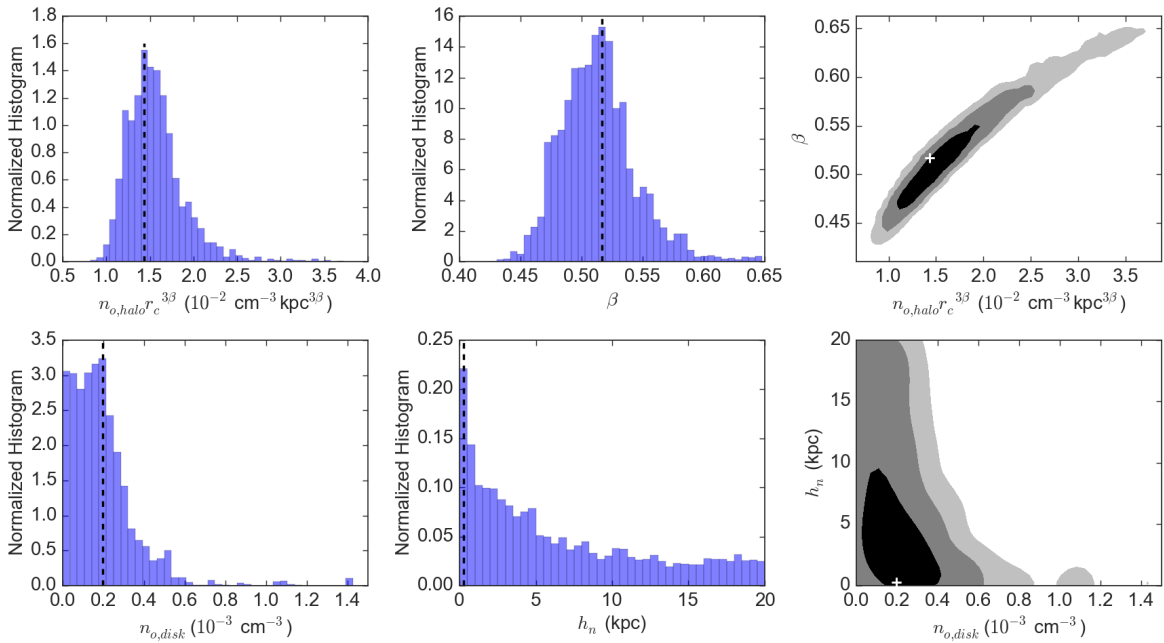}
\caption{Our MCMC fitting results as marginalized posterior probability distributions (left and center columns) and contour plots for the model parameters (right column).  The vertical dashed lines (distribution plots) and white crosses (contour plots) represent the best-fit model values.  The halo parameter distributions (top left and top center panels) are almost identical to the results from \cite{miller15}.  The disk midplane density distribution (bottom left panel) implies densities significantly lower than previous estimates ($\approx$10$^{-3}$ cm$^{-3}$).
}
\label{figure.disk_halo}
\end{center}
\end{figure*}

The Sun exists in a region known
as the Local Bubble, which emits soft X-rays
\citep[e.g.,][]{cox87,snowden97}, and we adopt for the Bubble a
temperature of $1.2\times 10^6$\,K, a variable path length between
100-300\,pc, and a density
$n_{\text{LB}}=4\times 10^{-3}$\,cm$^{-3}$. The nature of the Bubble and the
density remain debatable because of the contribution from solar-wind
charge exchange \citep[e.g.,][]{welsh09,kuntz09,galeazzi14,snowden15,henley15}. 
We have ignored hot interstellar gas confined within the galaxy disk other than
the Local Bubble.
Constraints from the intensity of the soft X-ray background imply that our
sight lines will incur only a small column from this material, and the
emission-line sample excludes the Galactic plane.
For the extended halo component, we
assume an isothermal ($2\times 10^6$\,K) plasma with a density profile
described by Equation~\ref{equation.beta_model}.
For the exponential disk component, we assume a form for the density of
\begin{equation}
n = n_{0,\text{disk}}e^{-z/h} \text{ cm$^{-3}$}
\label{equation.disk}
\end{equation}
where $h$ is the scale
height. The temperature declines in an analogous way, and we fix the midplane
temperature and its scale height to $2.5\times
10^6$\,K and 5.6\,kpc, respectively \citep{hagihara10} because we do
not model the line ratios as a temperature diagnostic. The free
parameters are the normalizations, $\beta$, and $h$, which we
constrain using the same MCMC method as \citet{miller15}.

We assume that each component is optically thin and compute its
contribution to the intensity along each line-of-sight $s$: 
\begin{equation}
I = \frac{1}{4\pi}\int n(s)^2 \epsilon(T(s)) ds
\end{equation}
where $\epsilon(T)$ is the volumetric line emissivity from the APEC
thermal plasma code \citep{foster12}. Finally, the line intensities
from \citet{henley12} do not account for absorption due to Galactic
H{\sc i}, so to compare our model intensities to theirs we apply
photoelectric absorption using the neutral hydrogen column from 
\citet{kalberla07} to the halo and disk components:
\begin{equation}
I(s) =
I_{\text{LB}}(s)+(I_{\text{halo}}(s)+I_{\text{disk}}(s))e^{-\sigma N_{\text{H}}}
\end{equation}
where $\sigma$ is the absorption cross-section and $I_{\text{LB}}$ is the
intensity from the Local Bubble. 

Figure~\ref{figure.disk_halo} shows the results from the MCMC analysis
as marginalized posterior probability distributions for each of the
free parameters. The model is not a significant improvement on the pure halo
model, and the best-fit parameters for the halo
($n_{0,\text{halo}}r_c^{3\beta} = 1.43\pm0.25 \times
10^{-2}$\,cm$^{-3}$\,kpc$^{3\beta}$, $\beta=0.52\pm0.03$) are
consistent with the results in \citet{miller15}. 
The disk parameters are poorly
constrained and indicate that \textit{the exponential disk contributes at most 10\% of the
total \ion{O}{7} column density.} This corroborates the \citet{miller15}
result. In contrast, the \citet{hagihara10} disk model (which assumes the
halo gas can be described entirely by a disk and is based on one sightline) 
finds $n_{0,\text{disk}} = 1.4\times 10^{-3}$
and $h=2.3$\,kpc. \citet{yao05}, who include some sight lines
near the Galactic plane and some outside the plane, find that the \ion{O}{7}
column is consistent with $n_{0,\text{disk}} =6.4\times 10^{-3}$\,cm$^{-3}$
and a scale height of $h=1.2$\,kpc, but the authors acknowledge that mixing
results from sight lines towards nearby X-ray binaries at low latitudes and
those at high latitudes can lead to a strong bias; the hot material confined
to the galaxy disk is not part of the exponential disk structure that we
have modeled, but it can contribute at low latitudes.

Thus, the impact of the disk component on the measured centroids must be
small, since 80-90\% of the column density will come from the 
spherical component. Assuming that the two components are kinematically
distinct, for the latter to be stationary and consistent with the
measured centroids requires a disk
speed much faster than co-rotation. We verified this by modeling a rotating
disk and a stationary halo where the disk contributes 10\% of the column. 
Since this is inconsistent with models where the gas originates in the galaxy
disk, the data probe motion in the spherical component. For reference, if
the \citet{hagihara10} disk model is adopted instead of the \citet{miller15}
model or our disk$+$halo model, the best-fit azimuthal
and radial velocities are
$v_{\phi} = 151\pm32$\,km\,s$^{-1}$ and $v_r = -15\pm18$\,km\,s$^{-1}$. 

We have assumed that, to first order, the spherical component moves as a solid
body with some $v_{\phi}$ and $v_r$. There may be second-order effects such as the internal flows
mentioned above (perhaps due to satellite motions or infalling clouds) or modes 
in the fluid, which would add scatter to the velocity measurements. However, 
if the halo is volume filling and in steady state, then large scale disturbances
will tend to dissipate in a sound-crossing time
(which could be a long time if the halo is very extended),
and the gas (if stationary)
will tend towards hydrostatic equilibrium. Multiple major kinematic components
are therefore not expected, although we would expect differential rotation. 
In this case, the measured rotation velocity is some average but still provides
useful information. Also, if the material is fresh infall from
the cosmic web then it likely accretes along filaments, which lead to a 
preferential orbital angular momentum axis. Finally, gas ejected from the disk
(either cold or hot) could spin up the halo \citep{marinacci11}, although
perhaps not to the velocity that we infer.

It is possible that there is a layer of warm-hot gas near the Galactic plane
(in addition to the Local Bubble) that affects every sight line out of the
Galaxy, but to which our models are not sensitive because we ignore data close
to the Galactic plane (where dust, supernova remnants, and other features make
modeling very difficult). Characterizing
the structure and filling factor of the hot interstellar medium has been a major effort
by itself \citep[e.g.,][]{yao05,nicastro16,hagihara11} and is beyond the scope of 
this paper, but if there is such a layer (possibly a disk with a scale height
of a few hundred pc) with a high density, the rotation signature in the RGS
data could be misattributed to the halo. On the other hand, for plausible
path lengths, oxygen column densities, and foreground absorption column
densities this layer should also produce emission in
excess of the Local Bubble contribution to the soft X-ray background.

To summarize, the halo models that are based on many data points as opposed to
a few sightlines favor a spherical halo (especially in emission) as the 
dominant component. 
Any contribution to the column density from a kinematically 
distinct (thick) exponential disk is small in this scenario,
so the Doppler shifts 
support a non-stationary extended halo. Reinterpreting these shifts will be necessary
if future data or analyses rule out a spherical halo (or at least a volume-filling
one) or a
high resolution X-ray spectrometer resolves the lines into components 
inconsistent with rotation. Also, we reiterate that the parametric fit strongly
depends on the few AGNs with the highest $S/N$, but the nonparametric tests
indicate rotation at some level.

\subsection{Potential Bias from Cooler Gas}

The \ion{O}{7} ion fraction is high between $3-20\times 10^5$\,K, so the cooler
(non-coronal) gas seen in and around the Galaxy in \ion{O}{6} with \textit{FUSE}  
\citep{nicastro03,savage03,wakker03,wang05,yao07,shelton07,bowen08} will
also absorb \ion{O}{7}. The \ion{O}{6} lines towards
background objects reveal Galactic and high-velocity absorbers \citep{nicastro03,savage03,sembach03}.
The Galactic absorbers (also seen towards stars) are consistent with an
exponential disk of scale height 1-4\,kpc \citep{savage03,bowen08}, whereas
the high-velocity absorbers may come from the halo or outside the Galaxy. 
If the column densities of these \ion{O}{6} absorbers are high, then their
\ion{O}{7} lines could bias our results. Since the disk component will rotate,
we are most concerned about contamination from this gas.

\citet{bowen08} measured \ion{O}{6} column densities for about 150
sight lines around the Galaxy and found a typical column of
$N_{\text{OVI}}\sin(b) \sim 1.6\times 10^{14}$\,cm$^{-2}$. For
the sight lines in our sample, we expect $N_{\text{OVI}} = 1-4 \times
10^{14}$\,cm$^{-2}$. If the \ion{O}{6} absorbers are at $T = 3\times
10^5$\,K and in collisional ionization equilibrium, the O{\sc
  vii}/\ion{O}{6} ratio is 1.7, which leads to an expected contribution
of $N_{\text{OVII}} = 2-7\times 10^{14}$\,cm$^{-2}$. In
contrast, the typical \ion{O}{7} column (assuming an optically thin
plasma) is $N_{\text{OVII}} \sim 4-5\times 10^{15}$\,cm$^{-2}$
\citep{miller13}. Hence, the contribution directly from \ion{O}{6} absorbers
in the Galaxy is at most 15\%. It is likely lower, since \citet{gupta12} and \citet{miller15}
argue that the \ion{O}{7} lines are not optically thin. The bias from the 
\ion{O}{6} absorbers also declines at higher latitude, since the \ion{O}{7}
column declines more slowly than the \ion{O}{6} column with increasing $b$
\citep[corroborated in \textit{emission} by][]{henley13}. For most of our
sight lines we estimate that the local and disk contribution to the \ion{O}{7}
column is less than 10\%. 

Even a 10\% contribution is important for measuring the true halo velocity,
but compared to the uncertainty in our best-fit velocity parameters, it is
small. More importantly, as with the hot disk considered above, it cannot by 
itself account for the measured velocities if we assume that the million-degree 
halo is stationary. 

\subsection{Summary and Conclusions}

We have measured a signature of rotation in the \ion{O}{7} line around
the Milky Way using 37 sight lines towards background AGNs and
archival RGS data. The parametric fit strongly depends on the
brightest AGNs, but nonparametric tests indicate that the \ion{O}{7}
absorbers are not stationary, even when removing suspect lines. 
From larger samples of emission and
absorption lines, we believe that an extended halo is more consistent
with the data than an exponential (thick) disk, and in this case this
halo rotates at some velocity smaller than $v_{\text{LSR}}$. Taken at
face value, this implies that the million-degree gas has a comparable
angular momentum to the galaxy disk.
It is possible that both components exist, in which case the RGS lines are
blends of kinematically distinct components. We use a large sample of
emission lines to constrain the contribution of each component to the column
density, and find that an exponential disk accounts for no more than 10\% of the
\ion{O}{7} column density. Thus, the measured Doppler shifts are dominated by
the motion of the gas in the extended halo, which is consistent with 
prograde rotation. Even if a disk were to dominate (which, for our data,
produces an unacceptable $\chi^2$ value), the best-fit
azimuthal velocities imply that it is rotating at nearly the same speed and
with a comparable amount of angular momentum to the spherical model.

This conclusion depends on assumptions about the underlying model. 
Measuring
the true velocity and separating the X-ray absorbers into their
various components requires a high-resolution X-ray instrument with a
large effective area \citep[e.g. \textit{ARCUS},][]{smith14}, which
would also give important information on their line shapes that could
constrain the optical depth and Doppler $b$ parameters 
\citep{miller16}, 
as well as reveal the contribution to the total column from the disk
and the coronal halo gas seen in X-ray emission.

The recent work on calibrating the conversion of dispersion angle to
a wavelength grid for the RGS \citep{devries15} and stacking multiple 
observations of the same object enables a wavelength accuracy of tens of
km\,s$^{-1}$ for the first time (Figure~\ref{figure.capella_wavelength}). After
investigating a variety of systematic issues, we find that the statistical
uncertainty (due to the low $S/N$ in many spectra and the relatively small
sample) remains the major source of error. The inferred halo velocity and
angular momentum are strongly model dependent (and the uncertainties that we
report are large), but the basic conclusion that the hot gas distribution 
rotates is less so.

Several scenarios could produce a rotating halo that lags behind the
disk, depending on the geometry. These include a galactic fountain of 
cool gas that spins up hot gas \citep{marinacci11}, a hot galactic fountain of
superbubble ejecta that produces an exponential disk before cooling,
or infall from the cosmic web with some preferential direction. Our
measurements cannot, by themselves, distinguish between these models
(which may not be mutually exclusive), but they are an important
kinematic constraint for future halo and galaxy formation models.

\acknowledgments

We thank Eric Bell for analysis suggestions and constructive
criticism, and Oleg Gnedin for information on the Galaxy's angular
momentum. Frits Paerels provided technical advice regarding the RGS. We
thank the referees for substantial comments that significantly
improved the manuscript.
We gratefully acknowledge financial support from the NASA ADAP
program, through grant NNX11AG55G.


\end{document}